\documentclass[usenatbib]{mnras}

\usepackage[T1]{fontenc}
\usepackage{ae,aecompl}
\usepackage{graphicx}
\usepackage{amsmath}
\usepackage{amssymb}
\usepackage{sidecap}
\usepackage{txfonts}
\usepackage{xcolor}
\usepackage{subfig}
\usepackage{appendix}
\usepackage{rotating}
\usepackage{hyperref}
\usepackage{upgreek}

\title[RX J1720.1+2638 with LOFAR at 54 MHz]{The ultra-steep diffuse radio emission observed in the cool-core cluster RX J1720.1+2638 with LOFAR at 54 MHz}

\pubyear{2021}

\begin{document}

\author[N. Biava et al.]{N. Biava,$^{1,2}$ F. de Gasperin,$^3$ A. Bonafede,$^{1,4}$ H. W. Edler, $^{3}$, S. Giacintucci, $^{5}$ \newauthor P. Mazzotta, $^{6}$ G. Brunetti, $^{4}$ A. Botteon, $^{7}$ M. Brüggen, $^{3}$ R. Cassano, $^{4}$  \newauthor A. Drabent, $^{8}$ A.C. Edge, $^{9}$ T. En{\ss}lin, $^{10}$ F. Gastaldello, $^{11}$ C.J. Riseley, $^{1,4,12}$ M. Rossetti $^{11}$  \newauthor H.J.A. Rottgering,$^{8}$ T.W. Shimwell, $^{13,8}$ C. Tasse, $^{14,15}$, R.J. van Weeren, $^{8}$   \\\\
$^1$ Dipartimento di Fisica e Astronomia, Università di Bologna, via P. Gobetti 93/2, I-40129, Bologna, Italy \\
$^2$ OAS-Bologna, INAF, via Gobetti 93/3, 40129 Bologna, Italy \\
$^3$ University of Hamburg, Gojenbergsweg 112, 21029 Hamburg, Germany \\ 
$^4$ IRA-Bologna, INAF, via P. Gobetti 101, 40129, Bologna, Italy \\
$^5$ Naval Research Laboratory, 4555 Overlook Avenue SW, Code 7213, Washington, DC 20375, USA \\
$^6$ Dipartimento di Fisica, Universit\`a di Roma "Tor Vergata", Via Della Ricerca Scientifica, 1, I-00133, Roma, Italy\\
$^7$ Leiden Observatory, Leiden University, PO Box 9513, 2300 RA Leiden, The Netherlands \\
$^8$ Th\"uringer Landessternwarte, Sternwarte 5, D-07778 Tautenburg, Germany\\
$^9$  Centre for Extragalactic Astronomy, Durham University, DH1 3LE, UK\\
$^{10}$ Max Planck Institute for Astrophysics, Karl-Schwarzschild-Str. 1, 85748 Garching, Germany\\
$^{11}$ IASF-Milano, INAF, via A. Corti 12, 20133, Milano, Italy \\
$^{12}$ CSIRO Space \& Astronomy, PO Box 1130, Bentley, WA 6102, Australia \\
$^{13}$ ASTRON, Netherlands Institute for Radio Astronomy, Oude Hoogeveensedijk 4, 7991 PD, Dwingeloo, The Netherlands \\
$^{14}$ GEPI \& USN, Observatoire de Paris, CNRS, Universite Paris Diderot, 5 place Jules Janssen, 92190 Meudon, France \\
$^{15}$ Department of Physics \& Electronics, Rhodes University, PO Box 94, Grahamstown, 6140, South Africa
}

\label{firstpage}
\pagerange{\pageref{firstpage}--\pageref{lastpage}}
\maketitle

\begin{abstract}
Diffuse radio emission at the centre of galaxy clusters has been observed both in merging clusters on scales of Mpc, called giant radio haloes, and in relaxed systems with a cool-core on smaller scales, named mini haloes. Giant radio haloes and mini haloes are thought to be distinct classes of sources. However, recent observations have revealed the presence of diffuse radio emission on Mpc scales in clusters that do not show strong dynamical activity. 
RX J1720.1+2638 is a cool-core cluster, presenting both a bright central mini halo and a fainter diffuse, steep-spectrum emission extending beyond the cluster core that resembles giant radio halo emission. In this paper, we present new observations performed with the LOFAR Low Band Antennas (LBA) at 54 MHz. These observations, combined with data at higher frequencies, allow us to constrain the spectral properties of the radio emission. The large-scale emission presents an ultra-steep spectrum with $\alpha_{54}^{144}\sim3.2$. The radio emission inside and outside the cluster core have strictly different properties, as there is a net change in spectral index and they follow different radio-X-ray surface brightness correlations. 
We argue that the large-scale diffuse emission is generated by particles re-acceleration after a minor merger.
While for the central mini halo we suggest that it could be generated by secondary electrons and positrons from hadronic interactions of relativistic nuclei with the dense cool-core gas, as an alternative to re-acceleration models. 

\end{abstract}


\section{Introduction}

Diffuse radio emission, not directly connected with single galaxies, has been observed in galaxy clusters, revealing the presence of large-scale magnetic fields and a non-thermal component of cosmic ray electrons (CRe) throughout the intracluster medium (ICM).
If this emission is located at the cluster centre, it is classified as giant radio haloes or mini haloes \citep[see e.g. review by][]{Brunetti2014,vanWeeren2019}.
Giant radio haloes are Mpc-sized sources predominantly found in massive, merging clusters \citep[e.g.][]{Cassano2010}. They show a steep radio spectrum, with a spectral index usually in the range $1.1 \le \alpha \le 1.4$ (where $S_{\nu} \propto \nu^{-\alpha}$).
The most commonly accepted scenario is that giant radio haloes are powered by re-acceleration of electrons by turbulence injected in the ICM during major merger events \citep[][]{Brunetti2001,Petrosian2001}.
Mini haloes are radio sources with size typically $\le 0.2\ R_{\rm 500}$ \citep[][]{Giacintucci2017,Giacintucci2019}, found in relaxed clusters with a cool-core, i.e. a core characterised by a peaked X-ray surface brightness and a significant drop in temperature ($T\le10^7 - 10^8$ K) at the centre.
The origin of mini haloes is still unclear. Two possible scenarios have been proposed: the hadronic and the turbulent re-acceleration models.
The hadronic model suggests that mini haloes are formed by the continuous injection of secondary electrons in the central regions by inelastic collisions of relativistic cosmic-ray protons with the cluster thermal proton population \citep{Pfrommer2004}. 
Another possibility is the re-acceleration of seed electrons in cluster cores by turbulence connected to the sloshing of the gas in the central potential well in response to a gravitational perturbation by a minor/off-axis merger \citep{ZuHone2013}.  
Mini haloes emission is often bounded by X-ray cold fronts, arc-shaped gas density discontinuities associated to gas sloshing \citep[][]{Mazzotta2008}.
The central AGN is a probable source of the seed electrons that are re-accelerated, and/or of the cosmic-ray protons that generate secondary electrons.
The hadronic model predicts a rather uniform spectrum for mini haloes with $\alpha\sim1$, while turbulent re-acceleration scenario is consistent with even steeper spectra \citep[][]{ZuHone2013,ZuHone2015}.
Recent findings of point-to-point X-ray and radio surface brightness comparisons show a super-linear behaviour, indicating a concentration of the ICM non-thermal components around the central AGN \citep[][]{Ignesti2020}.

Recent low-frequency observations with LOw Frequency ARray \citep[LOFAR;][]{vanHaarlem2013} High Band Antenna (HBA) at 144 MHz revealed the presence of diffuse radio emission on Mpc scales in two clusters that are relaxed and show no signs of major mergers \citep{Savini2019}. Up to now, only few such cases are
known in the literature \citep[][]{Bonafede2014, Venturi2017, Savini2018,Savini2019,Raja2020}. These sources, bridging the strict distinction between mini haloes and giant radio haloes, challenge our understanding of particle acceleration in the ICM. 
A possible explanation for the large-scale emission observed in these clusters is that minor mergers can produce enough turbulence in the ICM to re-accelerate particles on larger scales, without disrupting the cool-core.
If this is the case, the spectrum of the diffuse emission outside the core should be very steep ($\alpha\ge1.5$), as minor mergers are less energetic than major mergers \citep[][]{Cassano2006,Brunetti2008,Cassano2012,Brunetti2014}.
Hence, to understand the origin of these sources, the spectral properties and the spectral index distribution of their emission must be measured. 
However, due to observational limitations it has to-date been impossible to calculate the spectral index of the large-scale diffuse emission -- it has only been possible to place limits on the spectral index.

\begin{figure}
\centering
\includegraphics[width=0.5\textwidth]{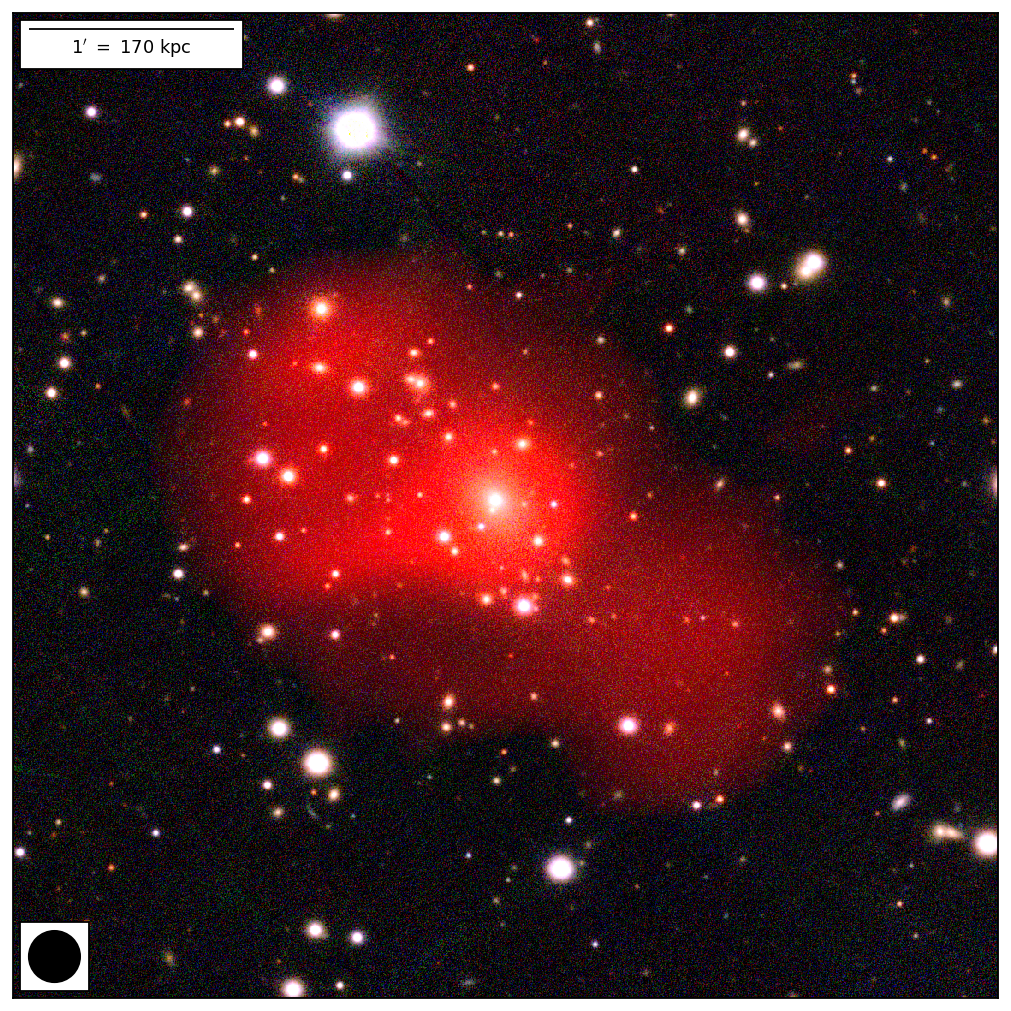}
\caption{Composite image of RXJ1720.1, obtained combining the radio LOFAR LBA image at 15 arcsec resolution in red (see Section \ref{sec:LBA}) with optical RGB Pan-STARRS1 images \citep[g, r, i bands;][]{Chambers2016}}
\label{fig:composite}
\end{figure}

In this paper we present new radio observations performed with LOFAR Low Band Antenna (LBA), centred at 54 MHz, of one of these peculiar sources, RX J1720.1+2638 (hereafter RXJ1720.1). 
This cool-core galaxy cluster \citep[cooling time $t_{\rm cool}=2.36$ Gyr,][]{Giles2017} presents a central mini halo and a nearby head-tail radio source north-east of the cluster centre, associated with a cluster member galaxy \citep[][]{Owers2011}.
A detailed radio spectral study of the source was performed by \cite{Giacintucci2014} in the frequency range $237 - 8440$ MHz, using data from the Giant Metrewave Radio Telescope (GMRT) and the Very Large Array (VLA). The mini halo consists of a bright central part, with a spectral index of $\alpha_{237}^{4850}=1.0\pm0.1$ and a lower surface brightness arc-shaped extension to the south-east with a steeper spectral index of $\alpha_{237}^{4850}=1.4\pm0.1$.
Two cold fronts detected in the X-ray band with \emph{Chandra} appear to confine the mini halo, suggesting the origin of the mini halo may be connected with gas sloshing that creates the cold fronts \citep[][]{Mazzotta2001, Mazzotta2008}.
A multi-object optical spectroscopic study of this cluster identified two group-scale substructures which could have perturbed the cluster core \citep{Owers2011}. The most promising perturber candidate lies $\sim550$ kpc north of the cluster core, spatially coincident with a peak in the weak lensing maps of \cite{Okabe2010}.A weak enhancement in the lensing maps is also present roughly at the position of the substructure located $\sim400$ kpc to the south-west, which, however, is dynamically less significant.
LOFAR HBA (144 MHz) observations revealed a new diffuse component extending beyond the cluster core, south-west of this, with an overall size of $\sim600$ kpc \citep{Savini2019}. Since this emission is not visible at higher frequencies, combining LOFAR 144 MHz and GMRT 610 MHz data, \cite{Savini2019} could only provide a lower limit of the spectral index, of $\alpha\ge1.5$ in the vicinity of the mini halo up to $\alpha\ge2.1$ at larger distances. 
To constrain the spectral properties of the newly discovered emission, observations at lower frequencies are therefore mandatory.

The aim of this work is to calculate for the first time the spectral index of the large-scale diffuse radio emission.
Combining LOFAR LBA and HBA data we could create a spectral index map of the whole diffuse emission in this cluster.
To understand the origin of the diffuse radio emission inside and outside the cluster core, we also analysed archival radio and X-ray data, looking for a possible curvature in the radio emission and a spatial correlation between radio and X-ray surface brightness.
A composite image of the source, created with our new LOFAR LBA data, is shown in Fig. \ref{fig:composite}.

The paper is organized as follows: in Section \ref{sec:data} we describe the data used in our analysis and their reduction. The results of the radio data are presented in Section \ref{sec:results}. A comparison between radio and X-ray data is reported in Section \ref{sec:comparison}. In Section \ref{sec:discussion} we discuss our results and we present our conclusions in Section \ref{sec:conclusions}. 

The cluster RXJ1720.1 (RA: 17:20:10.1 DEC: +26:37:29.5, \cite{Piffaretti2011}) is located at redshift $z = 0.164$ \citep[][]{Harris1988, Crawford1999}. This corresponds to a scale of $2.8\ \rm{kpc\ arcsec^{-1}}$ (adopting a $\Lambda$CDM cosmological model with $\Omega_{\Lambda} = 0.7$, $\Omega_m = 0.3$ and $H_0 = 70\ {\rm km\ s^{-1}\ Mpc^{-1}}$).

\section{Data reduction}\label{sec:data}
In this paper we present new LOFAR LBA (54 MHz) observations of the galaxy cluster RXJ1720.1.
These observations are complemented by LOFAR HBA (144 MHz), GMRT (610 MHz) and VLA L-band (1480 MHz) and C-band (4860 MHz) observations.
In the following sections, we describe the data reduction procedures of LOFAR LBA and HBA and VLA observations. For GMRT data reduction instead we refer to \cite{Savini2019}.
All radio observations used in the paper are summarized in Table 1.

\begin{table*}
    \centering
    \caption{Summary of the observational details.}
    \renewcommand\arraystretch{1.2}
    \begin{tabular}{cccccccc} \hline
         Telescope &Central frequency &Configuration &Time on source &Observation ID &Observation date &Reference\\\hline
         LOFAR &54 MHz &\verb|LBA_OUTER| &10h &LC12\_018 &26 Sept, 20 Nov 2019 &1\\
         LOFAR &144 MHz &\verb|HBA_DUAL_INNER| &16h &LC7\_024 , LT10\_010 &25 Jan 2017, 6 Mar 2020 &1\\
         GMRT & 612 MHz &- &4h &11MOA01 &10 Mar 2007 &2\\
         VLA &1480 MHz &B &70min &AH190 &25 Apr 1985 &3\\
         VLA &4860 MHz &C &4min &AE125 &16 Jan  1999 &3\\
         \hline \end{tabular}
    \\References: (1) this article, (2) \cite{Savini2019}, (3) \cite{Giacintucci2014}.
        \label{tab:obs}
\end{table*}

\begin{figure*}
\centering
\includegraphics[width=\textwidth]{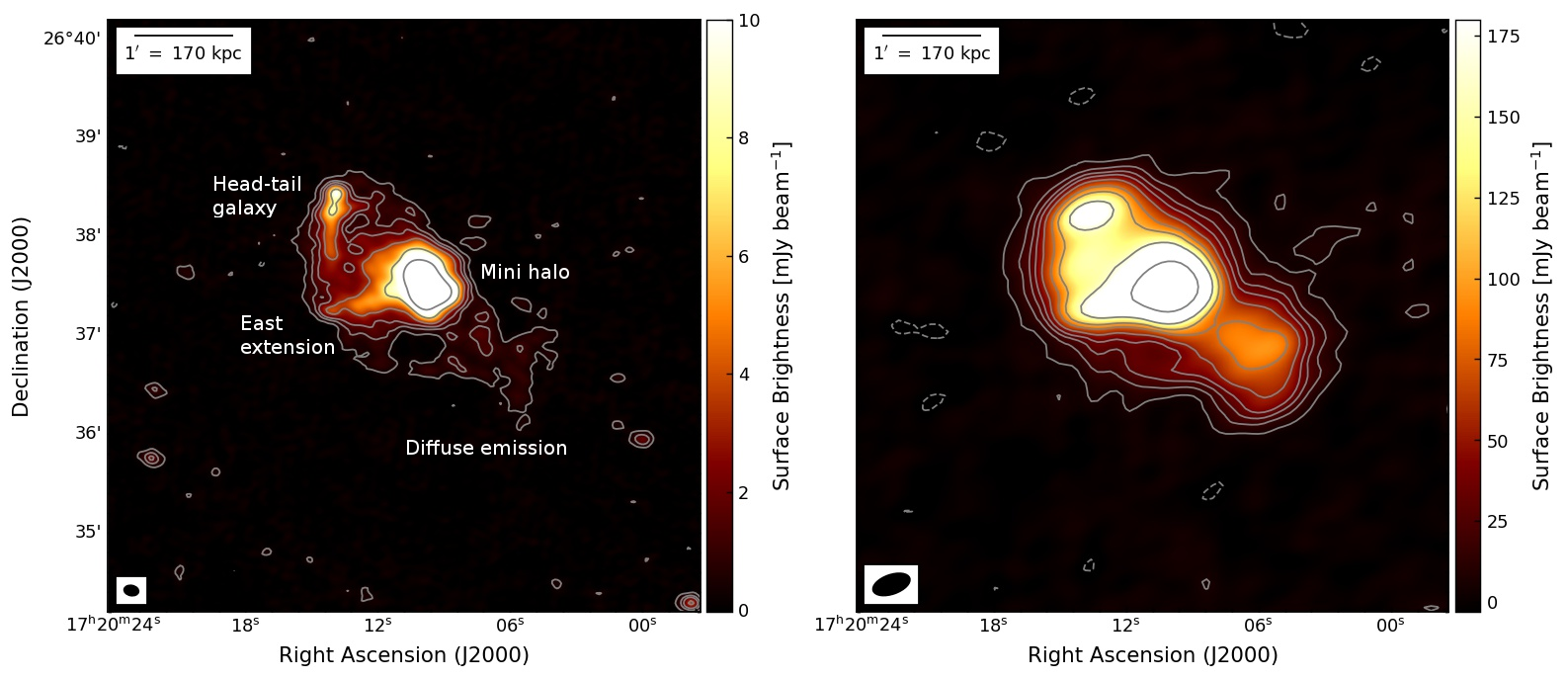}
\caption{Radio maps of RXJ1720.1 at different frequencies. \emph{Left:} LOFAR HBA 144 MHz at 9 arcsec $\times$ 7 arcsec resolution. Levels:[-1,1,3,5,9,20,50] $\times\ 3\sigma$ (where $\sigma=0.125\ \rm{mJy\ beam^{-1}}$). \emph{Right:} LOFAR LBA 54 MHz at 23 arcsec $\times$ 12 arcsec. Levels:[-1,1,3,5,7,10,15,20,30,70] $\times\ 3\sigma$ (where $\sigma=1.8\ \rm{mJy\ beam^{-1}}$). The beam is shown in the bottom left corner of each image.}
\label{fig:lofar}
\end{figure*}

\subsection{LOFAR LBA observations}\label{sec:LBA}
The galaxy cluster RXJ1720.1 was observed with LOFAR LBA on 2019 September 26 and November 20, for a total observing time of 10 h. The observations were performed in the frequency range $30-78$~MHz in a dual-beam mode, with one beam continuously pointing at the calibrator (3C295) and one beam at the target. We used the \verb|LBA_OUTER| antenna mode, where the outer 48 dipoles of the station are used to minimize the coupling between dipoles, using 24 Core Stations and 14 Remote Stations. 

The data were taken at 1s integration time and at a frequency resolution of 64 channels per 0.192 MHz sub band and subsequently averaged to 2 s and 16 ch/sb.
The calibrator data were reduced following \cite{deGasperin2019}, to isolate the systematic effects of the polarisation alignment, the bandpass and the clock drifts. Since in the low-frequency regime the last effect is difficult to isolate from the ionospheric delay, the raw scalar phase solutions are transferred to the target. This also gives an initial estimate of the ionospheric delay, using the solutions towards the direction of the calibrator. After the application of the calibrator solutions, the data for the target field required different steps of calibration to correct for differential ionospheric effects. For the target field calibration we followed the procedure described in \cite{deGasperin2020}. An initial directions independent (DI) calibration of the target field removed three systematic effects: the direction-averaged ionospheric delay, the Faraday rotation and beam variations with time and frequency on top of the LOFAR beam model.

The primary errors remaining in the data at this point
are the severe differential directions dependent (DD) effects caused by the ionosphere. To correct for these errors, bright sources in the field of view, called DD-calibrators, selected based on their flux density, are used to estimate the ionospheric effects in their directions. To obtain a high-fidelity image of RXJ1720.1, we adapt the selection criteria of DD-calibrators to include also our target. Then, the field of view is divided into facets given by the Voronoi-tesselation of the positions of the DD-calibrators. Each facet is corrected with the calibration solutions of the corresponding DD-calibrator during imaging with \verb|DDFacet| \citep[][]{Tasse2018}. As a result, we obtain a DD calibrated wide-field image at the full resolution of 15 arcsec.

The final data reduction step is the extraction and self-calibration of the target, where we employ a LBA-specific implementation of the extraction strategy described in \cite{vanweeren2020}.
We subtract all sources outside the region of interest from the full data set, using the model and the calibration solutions derived in the DD calibration. To further refine the image quality of the target, we perform scalar phase and slow diagonal amplitude self-calibration at increasing time-resolution. 
The flux density scale was set according to \cite{ScaifeHeald2012}. The flux calibration uncertainty is estimated to be 10 per cent \citep[][]{deGasperin2021}.
We imaged the data with \verb|WSClean| \citep[][]{Offringa2014,Offringa2017} using a Briggs weighting scheme with robust=$-0.5$, applying an inner $uv$-cut at $30\lambda$ and using a multi-scale deconvolution.
The final image of RXJ1720.1 has a resolution of 23 arcsec $\times$ 12 arcsec and a noise of $1.8\ \rm{mJy\ beam^{-1}}$
(see Fig. \ref{fig:lofar}, right panel).

\subsection{LOFAR HBA observations}
The LOFAR HBA data used in this work are part of the LOFAR Two Meter Sky Survey \citep[LoTSS, see][]{Shimwell2017}.
The first LOFAR HBA study of RXJ1720.1 was presented by \cite{Savini2019}. In that work, the authors used the LoTSS pointing P260+28 observed on 2017 January 25, and the data were calibrated using FACTOR \citep{Factor2016}.

Here, we use the same pointing along with a more recent one, P261+25 observed on 2020 March 6, both processed with the standard
Surveys Key Science Project pipeline \footnote{https://github.com/mhardcastle/ddf-pipeline/} \citep[see][]{Shimwell2019,Tasse2021}. 
Each LoTSS pointing consists of an 8 hr observation book-ended by
10 min scans of the flux density calibrator using HBA stations in
the \verb|HBA_DUAL_INNER| mode, implying that only central antennas are used for the remote stations to mimic the size of the core stations. 
The data were corrected for the directions independent effects following the procedure described in \cite{Shimwell2017} and the directions dependent effects through the pipeline that uses \verb|killMS| and \verb|DDF| \citep{Tasse2014a,Tasse2014b,SmirnovTasse2015,Tasse2018,Tasse2021} for direction dependent calibration and imaging respectively. 
To improve the accuracy of the calibration towards the target, an additional common self-calibration of
the two DD calibrated data sets was performed \citep[see][for details]{vanweeren2020}. After the subtraction of all sources outside a small region containing the target, using the DD gains, several iterations of phase and amplitude self-calibration in the extracted region were performed using \verb|DPPP| and \verb|WSclean| \citep{DPPP2018,Offringa2014,Offringa2017}. 
The flux density scale was set according to \cite{ScaifeHeald2012}, and subsequently aligned with LoTSS-DR2 data release, where the flux calibration uncertainty is estimated to be 10 per cent \citep[][Shimwell et al. in prep]{Hardcastle2021}.
The final high-resolution image was produced using a Briggs weighting scheme with robust =$ -0.5$, applying an inner $uv$-cut at $80\lambda$ and using a multi-scale deconvolution. The final image has a resolution of 8.8 arcsec $\times$ 6.1 arcsec and an rms of $0.125\ \rm{mJy\ beam^{-1}}$ (see Fig. \ref{fig:lofar}, left panel).

\subsection{VLA data reduction}
RXJ1720.1 was observed with the VLA in a number of different bands and configurations. We used the data at 1480 MHz, B configuration, and at 4860 MHz, C configuration.
The data sets were reduced with \verb|CASA| \citep[version 5.4,][]{McMullin2007} following standard procedures after manual flagging.
To improve the quality of the final image at 1480 MHz, we performed several cycles of phase self-calibration, to reduce the effects of residual phase variations in the data. Self-calibration was also attempted at 4860 MHz, but the solutions were not applied as the process did not improve the gain solutions due to the low flux density of the source. At both frequencies, the flux density scale was set according to VLA Perley (1990) values in SETJY task. The flux calibration uncertainty is estimated to be $5$ per cent at all frequencies.

\section{Results}\label{sec:results}

\subsection{Morphology}
In Fig. \ref{fig:lofar}, we show on the left the LOFAR HBA image at a central frequency of 144 MHz and with a resolution of 9 arcsec $\times$ 7 arcsec, while on the right the LOFAR LBA image at central frequency of 54 MHz and with a resolution of 23 arcsec $\times$ 12 arcsec.
Thanks to the high resolution of the HBA image we can identify several radio sources, indicated in the figure.
The central diffuse emission is composed of a bright mini halo with a size of $r\sim25\ \rm{arcsec}$ ($\sim70$ kpc), and a lower surface brightness extension to the east. 
Both components were already observed at 612 MHz by \cite{Giacintucci2014}. However, in their image the fainter emission has an arc-shaped morphology extending to the south-east. A reanalysis of the same data carried out by \cite{Savini2019} does not show this additional emission towards the south, recovering the morphology that we observe at 144 MHz.
North-east of cluster centre there is a head-tail radio galaxy, which appears connected to the mini halo by a faint emission region.
However, these two components could overlap for projection effects, without being physically connected.
To the south-west of the mini halo, diffuse emission on larger scales is visible. We note that this emission extends also to the north, and essentially surrounds the central mini halo.
This diffuse emission is brighter and more evident in the LBA image. Its extension is comparable in the two images ($\sim200$ arcsec $\cong$ 560 kpc) and is elongated in the north-east south-west direction.

\begin{figure*}
\subfloat{
\centering
\includegraphics[width=\textwidth]{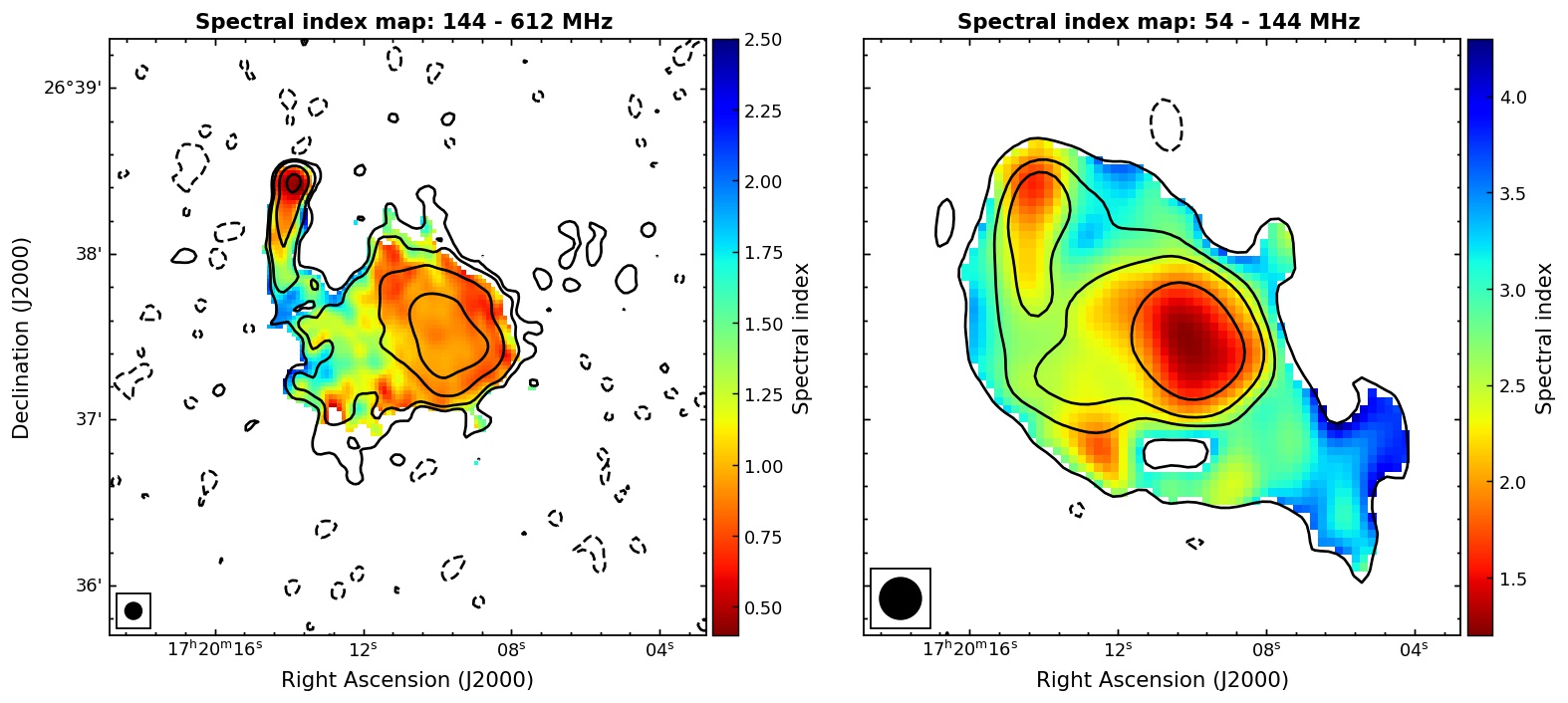}} \\
\subfloat{
\centering
\includegraphics[width=\textwidth]{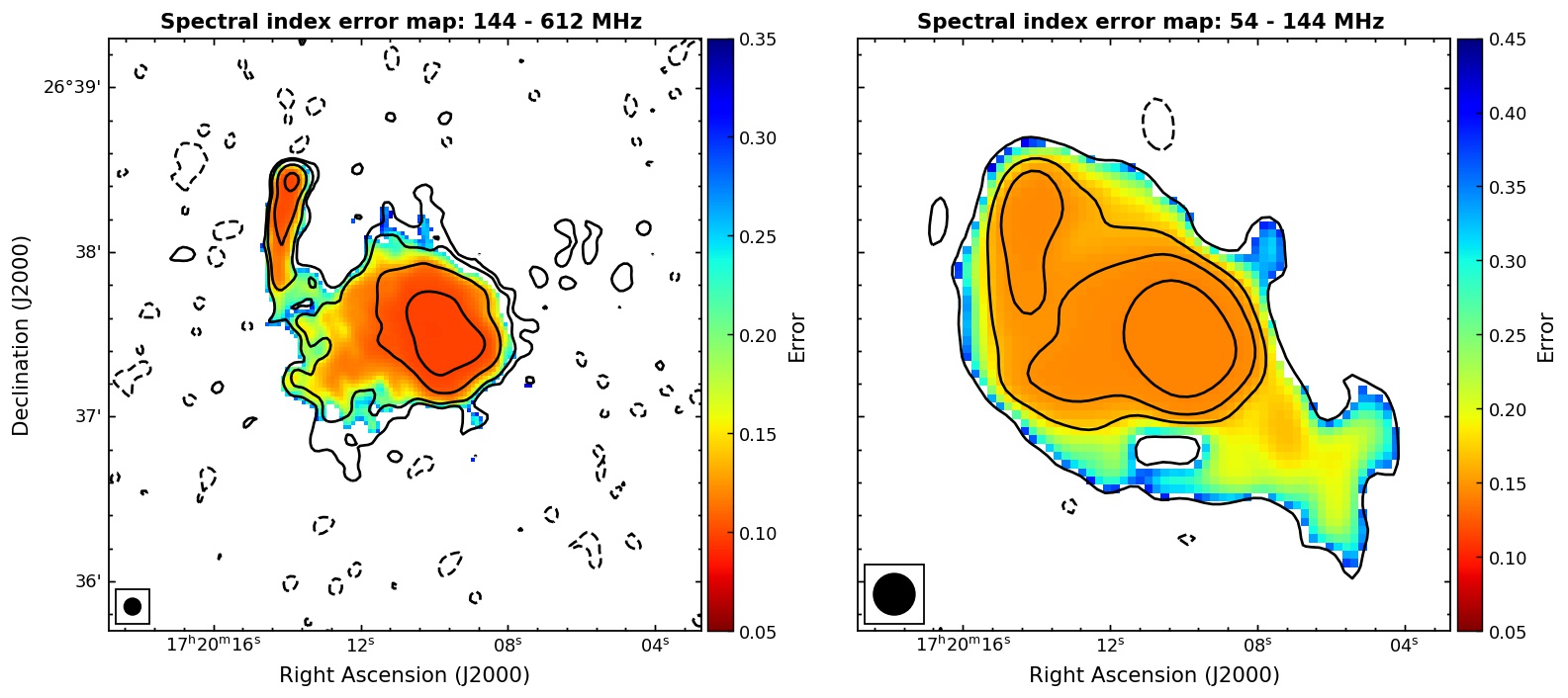}}
\caption{Spectral index maps (\emph{Top}) and associated spectral index error maps (\emph{Bottom}) of RXJ1720.1. \emph{Left}: frequency range 144 - 612 MHz, resolution 6 arcsec. Overlaid are the GMRT 612 MHz contours. \emph{Right:} frequency range 54 - 144 MHz, resolution 15 arcsec. Overlaid are the LOFAR HBA 144 MHz contours. The beam is shown in the bottom left corner of each map.}
\label{fig:spix}
\end{figure*}

\begin{figure*}
\centering
\includegraphics[width=\textwidth]{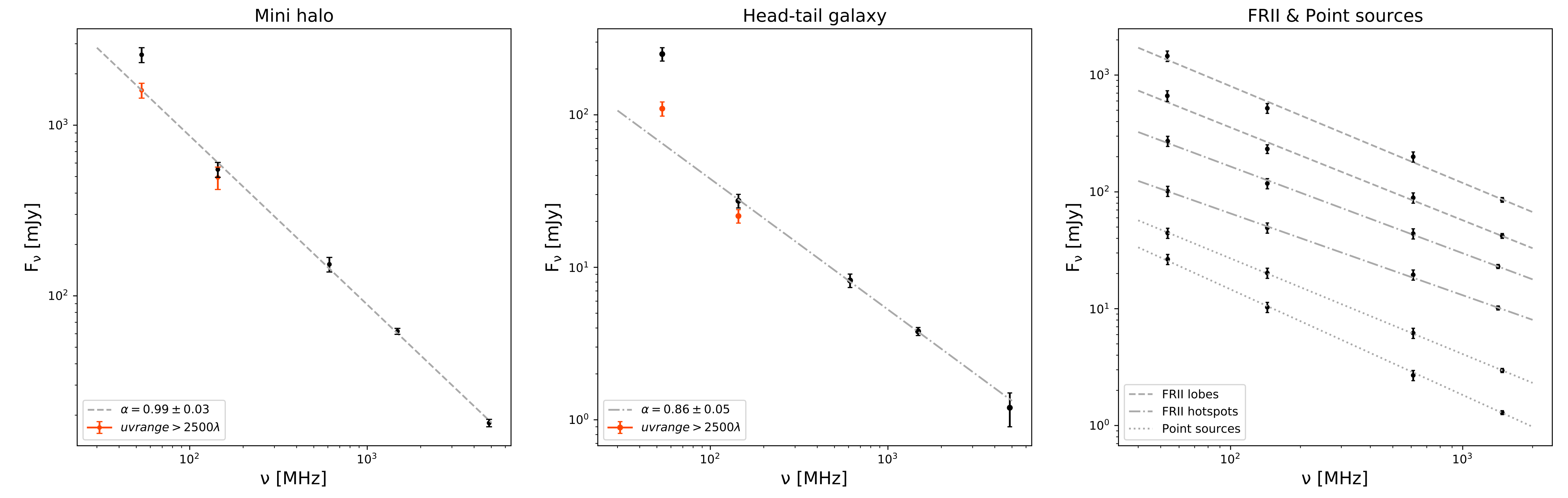}
\caption{\emph{Left:} Integrated flux of the mini halo from the original images (black points) and LOFAR images with $uv$-range > $2500\lambda$ (orange points). Dashed line obtained from the weighted least square of the black points. \emph{Middle}: Integrated flux of the core of the head-tail radio galaxy from the original images (black points) and LOFAR images with $uv$-range > $2500\lambda$ (orange points). Dashed-dotted line obtained from the weighted least square of the black points. \emph{Right:} Weighted least square of integrated flux of the FRII radio lobes (dashed lines) and hotspots (dashed-dotted lines) and of point sources located near the cluster (dotted lines).}
\label{fig:int_flux}
\end{figure*}

\subsection{Spectral analysis}
In order to understand whether we are observing two different types of emission (a central mini halo and a giant radio halo on larger scales), we investigate the spectral index distribution of the radio emission.
To study in detail the different components, we produced two spectral index maps: one at a resolution of 6 arcsec between LOFAR HBA (144 MHz) and GMRT (612 MHz) data (Fig. \ref{fig:spix}, left panels) and one at lower resolution of 15 arcsec between LOFAR LBA (54 MHz) and HBA data (Fig. \ref{fig:spix}, right panels).
To create these maps we re-imaged all the radio data with the same $uv$-range ($690\sim4000\ \lambda$) and using a uniform weighting scheme. Then we spatially aligned the images with the same restoring beam, to correct for possible shifts introduced by the phase self-calibration process. Finally, we considered only the emission detected above $3\sigma$ and we assumed a flux calibration error of 10 per cent for both LOFAR and GMRT data.

A previous spectral index map of the cluster between 144 and 612 MHz was presented by \cite{Savini2019} with a resolution of 20 arcsec and they found $\alpha\sim1$ for the mini halo and put a lower limit of $\alpha\ge 1.5$ for the diffuse emission outside the cluster core. 
Our increased resolution, sensitivity and frequency coverage allows us to study the spectrum in much more details. 
In our map between 144 and 612 MHz, the mini halo shows quite a uniform spectral index with a mean value of $\alpha_{144}^{612}=0.93\pm0.10$ and variations of $\Delta\alpha\sim0.1$ across the region . The east extension of the central emission has a steeper spectral index, with values between $\alpha_{144}^{612}=1.07\pm0.15$ and $\alpha_{144}^{612}=1.84\pm0.15$. This is in agreement with the trend observed by \cite{Giacintucci2014} at higher frequencies. 
The core of the head-tail radio galaxy in the north shows a typical spectral index $\alpha_{144}^{612}\sim0.6$ which steepens along the tail, indicative of ageing, as observed in radio galaxies.
To detect the diffuse emission outside the cluster core, we have to go to lower frequencies, in the range 54 $-$ 144 MHz.
This emission shows an ultra-steep spectrum, with a mean spectral index of $\alpha_{54}^{144}=3.2\pm0.2$.
This is the first cool-core cluster for which the spectral index of large-scale diffuse emission was estimated, and its large value is in agreement with the lower limit provided by \cite{Savini2018}. 

In the low-frequency spectral index map (Fig. \ref{fig:spix}, right panel) we noticed a net difference in the spectral index of the mini halo and the diffuse emission on larger scales. 
This suggests that the radio emission inside and outside the cluster core has a different nature. 
The strong spectral gradient observed (in projection) across the boundaries of the mini halo rules out the possibility that the electrons producing the emission on larger scale are advected/transported from the mini halo region on larger volumes. 

However, in this frequency range the spectral index of the mini halo and of the core of the head-tail radio galaxy is steeper than the one found at higher frequencies.
To confirm this behaviour, we created an image of the cluster at all available frequencies (VLA data included) with the same parameters ($uv$-range = $690\sim40000\ \lambda$, uniform weighting scheme, and 15 arcsec restoring beam) to extract the integrated flux from mini halo region and from the core of the head-tail radio galaxy. 
From the spectrum of the mini halo, plotted in Fig. \ref{fig:int_flux} left panel, we can see that the flux density at 54 MHz is larger than the expected value from a spectral index of about 1, as found in the frequency range 144 $-$ 4860 MHz (dashed line). 
In fact, the measured spectral index between 54 $-$ 144 MHz is $\alpha_{54}^{144}=1.57\pm0.14$, while at higher frequencies it is flatter ($\alpha_{144}^{612}=0.88\pm0.10$, $\alpha_{612}^{1480}=1.02\pm0.12$, $\alpha_{1480}^{4860}=1.04\pm0.05$).
A similar offset is observed for the head-tail radio galaxy (Fig. \ref{fig:int_flux}, middle panel), where $\alpha_{54}^{144}=2.24\pm0.14$, while $\alpha_{144}^{612}=0.83\pm0.10$, $\alpha_{612}^{1480}=0.87\pm0.13$ and $\alpha_{1480}^{4860}=0.97\pm0.22$.
The spectra of the core of an active galaxy is expected to follow a pure power-law with a spectral index around 0.6 $-$ 0.9. Then the spectrum becomes steeper in the high frequency range because of particles ageing.
Hence, the steep low-frequency spectral index we observe cannot be explained by the radio galaxy alone, and could be due to the superposition of components with different spectral behaviours.

To understand if there are offsets in the LBA flux density scale, we computed the spectrum of point sources in the proximity of the cluster and of one extended nearby FRII radio galaxy (NVSS J172104+262417). This allows us to check possible systematic errors in the absolute flux calibration, and possible errors introduced by the deconvolution of extended sources. In Fig. \ref{fig:int_flux} right panel, we show the spectrum of the point sources and of the radio lobes and the hotspots (comparable to the case of the head-tail radio galaxy in our target) of the extended radio galaxy.
For all control sources we do not observe particular offsets of the LBA fluxes, the points are consistent within the error with the higher frequency part of the spectrum, assuming these sources follow a power-law spectrum down to low frequencies.
Hence, we can conclude that the offsets observed in the mini halo and head-tail radio galaxy are not affected by systematic errors.
We argue that the higher flux density we observed in our target at LBA frequency, it is due to a superposition of two different components along the line of sight: the steep diffuse emission and the mini halo/head-tail core.
In support of this, we note that the LBA emission in the mini halo region is comparable to the sum between the flux extrapolated from higher frequencies and the flux of the steep large-scale diffuse emission, scaled for the region area, assuming it has the same brightness as in the regions outside the mini halo. 
There is therefore evidence that the steep emission observed south-west of the cluster core is substantially present also in the central regions, resembling a giant radio halo.
As a further check, we tried to isolate the mini halo flux from that of the steep diffuse emission, re-imaging the LOFAR data with a cut in the $uv$-range.
We selected a $uv$-range > 2500 $\lambda$ (corresponding to $\sim 190$ kpc), which allows us to somewhat filter out the larger scale steep component of the diffuse emission in the centre of the cluster, without losing the emission coming from the mini halo.
We found that with the imposed $uv$-range restriction, the LBA flux density of mini halo and head-tail radio galaxy decreases (Fig. \ref{fig:int_flux}, orange points in the left and middle panels), confirming our hypothesis. 

\begin{figure}
\centering
\includegraphics[width=0.5\textwidth]{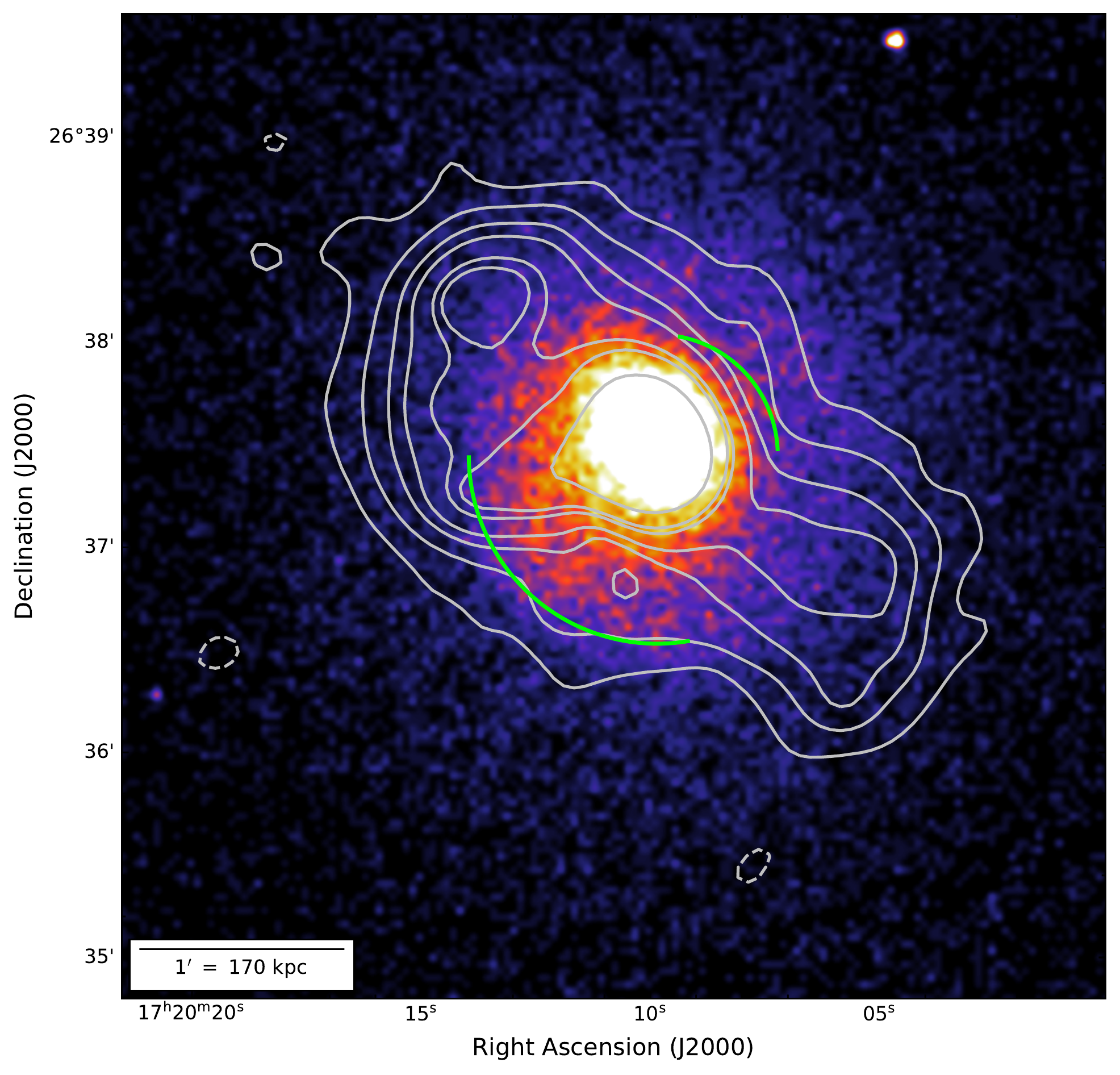}
\caption{\emph{Chandra} X-ray image of RXJ1720.1 in the 0.5 - 2.5 keV energy band. The background has been subtracted and the image has been corrected for the total exposure and the effective area of the telescope. Each pixel corresponds to 2 arcsec. Overlaid are the LOFAR LBA contours at 15 arcsec resolution. Green lines indicate the position of the cold fronts.}
\label{fig:X}
\end{figure}


\begin{figure*}
\subfloat{
\centering
\includegraphics[width=0.4\textwidth]{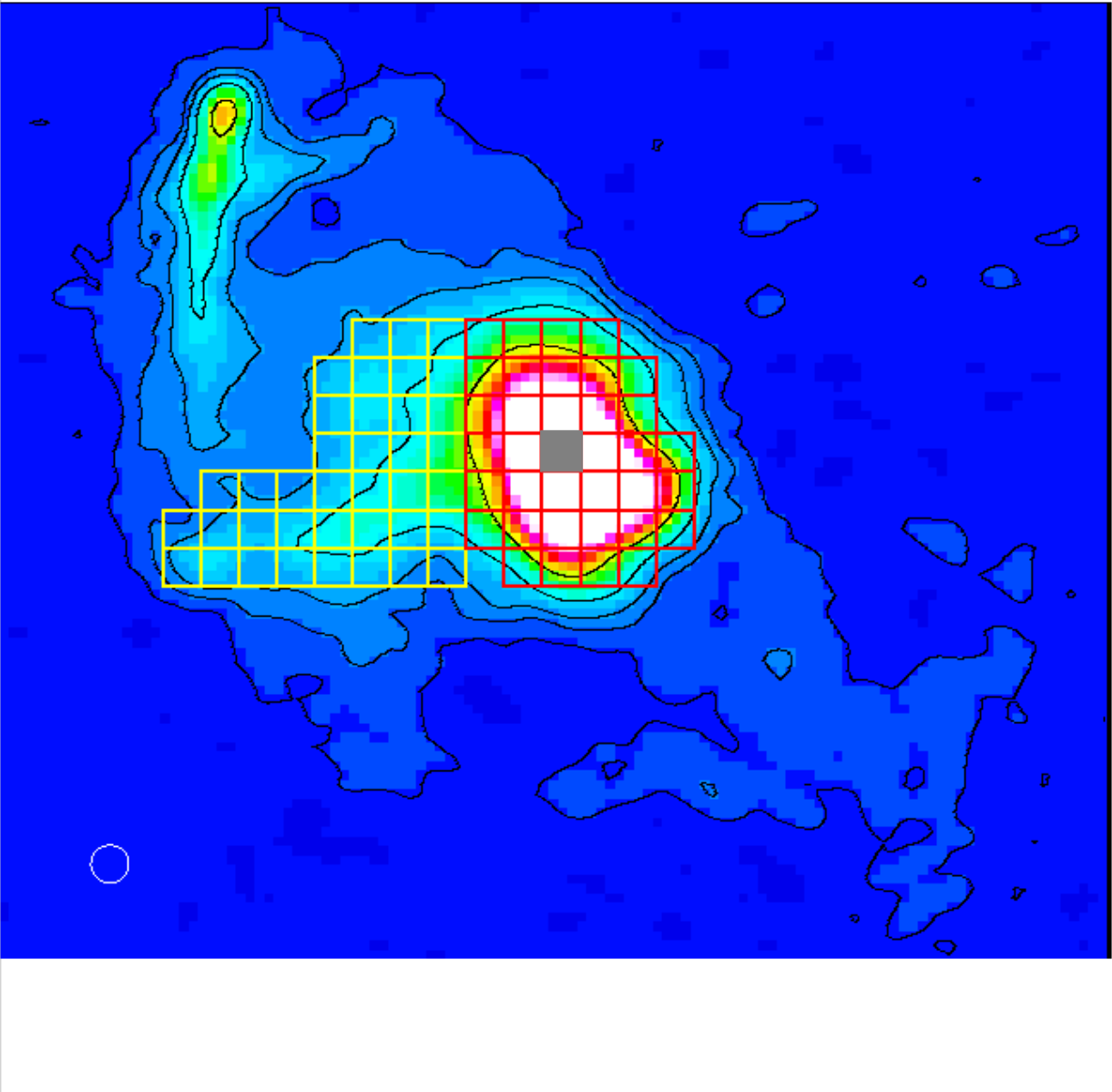}}
\qquad
\subfloat{
\includegraphics[width=0.5\textwidth]{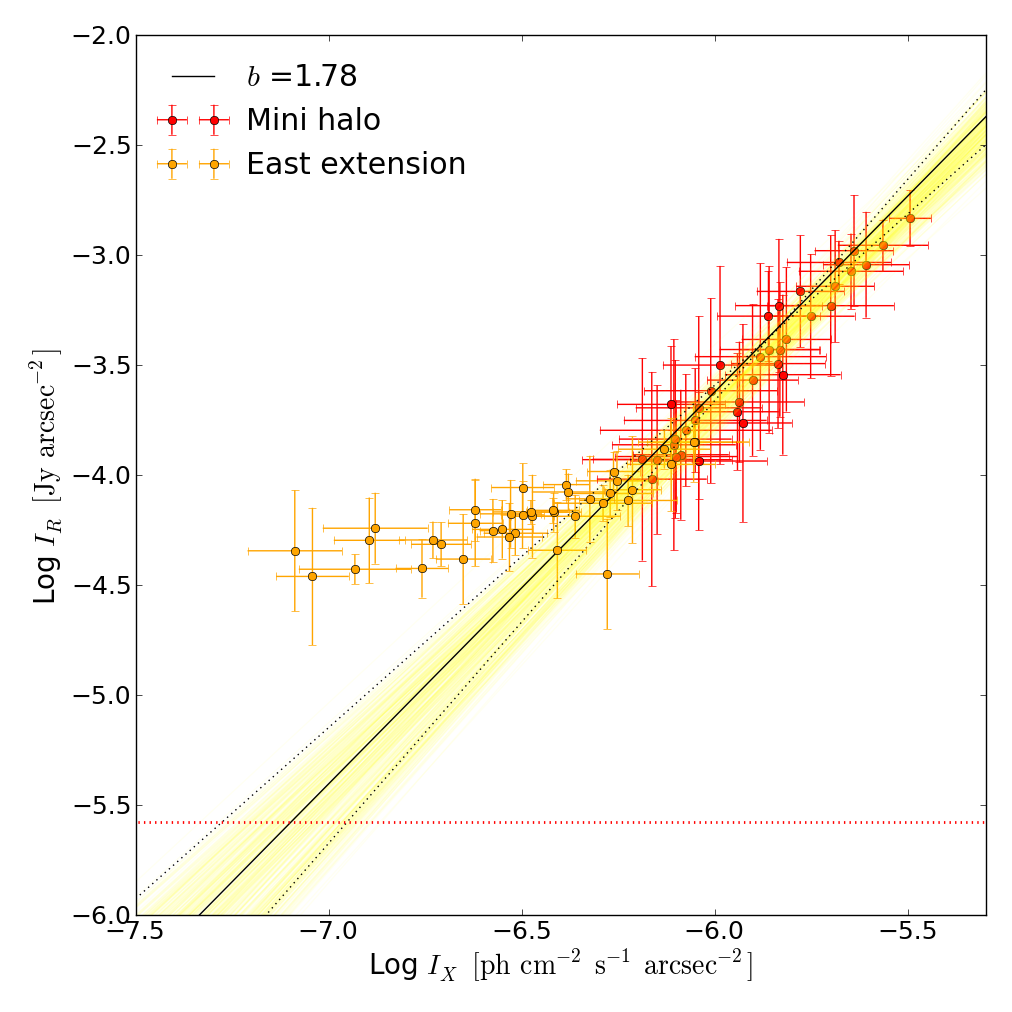}}
\caption{Radio-X-ray correlation for the mini halo. \emph{Left:} LOFAR HBA image of RXJ1720.1 at 6 arcsec resolution with contours at 3,9,15,30,60 $\times$ $\sigma$, where $\sigma=0.108\ \rm{mJy\ beam^{-1}}$. With overlaid the grid on the mini halo emission in red and on the east extension in yellow. The contribution of the bright central galaxy is excluded from the sampling (grey region). The circle in the lower left corner shows the beam. \emph{Right:} Radio-X-ray surface brightness correlation of the central emission. Red points, with associated statistical errors, are from the mini halo region, while orange points from the east extension. The black line shows the best-fit correlation on mini halo emission, black dotted lines show the 25th and 75th percentile of the posterior distribution. The red dotted line marks $1\sigma$.}
\label{fig:Corr_MH}
\end{figure*}

\begin{figure*}
\subfloat{
\centering
\includegraphics[width=0.5\textwidth]{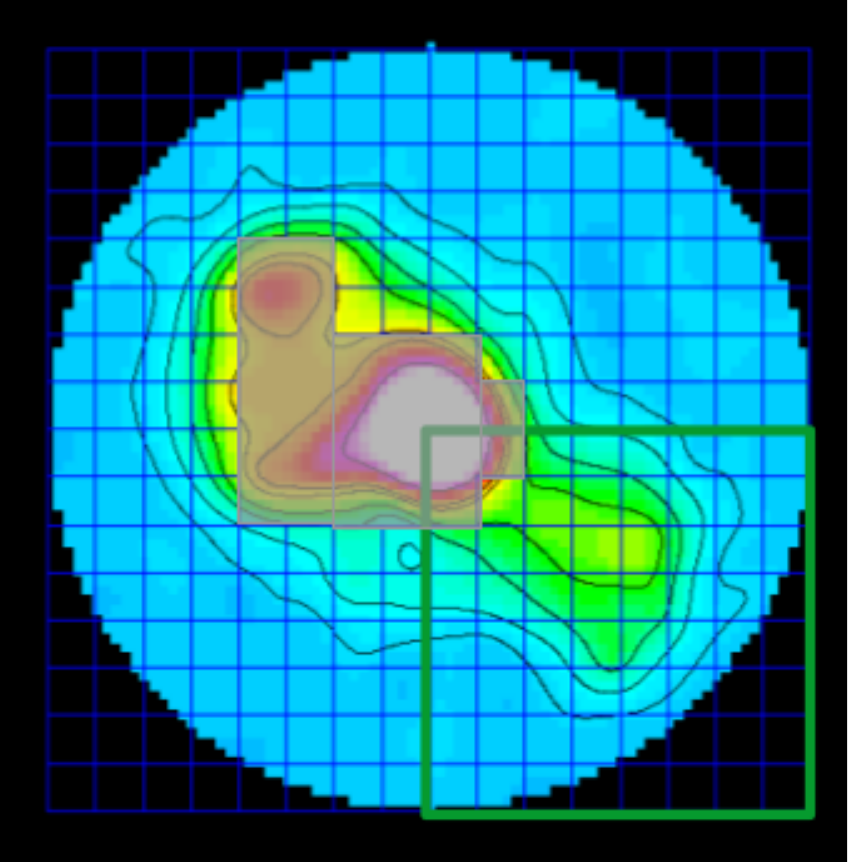}}

\subfloat{
\includegraphics[width=0.5\textwidth]{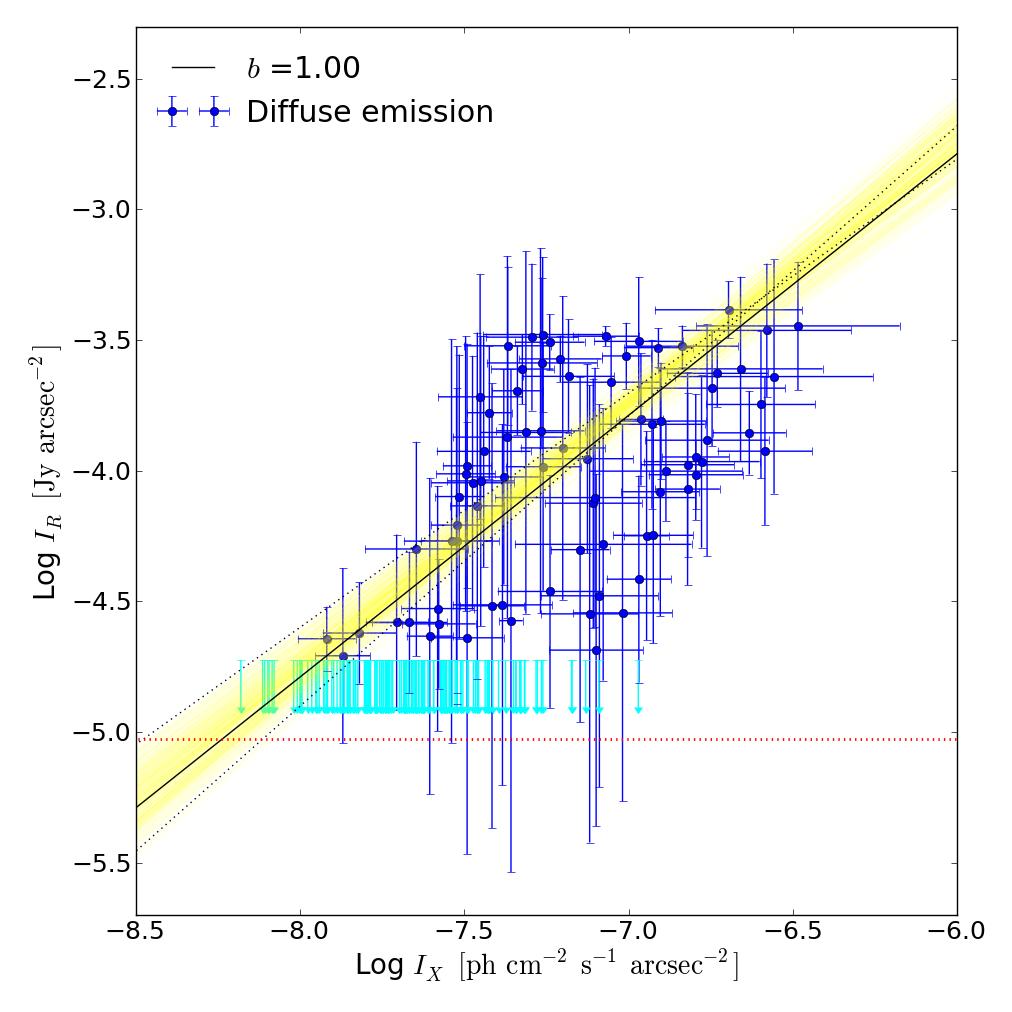}
\includegraphics[width=0.5\textwidth]{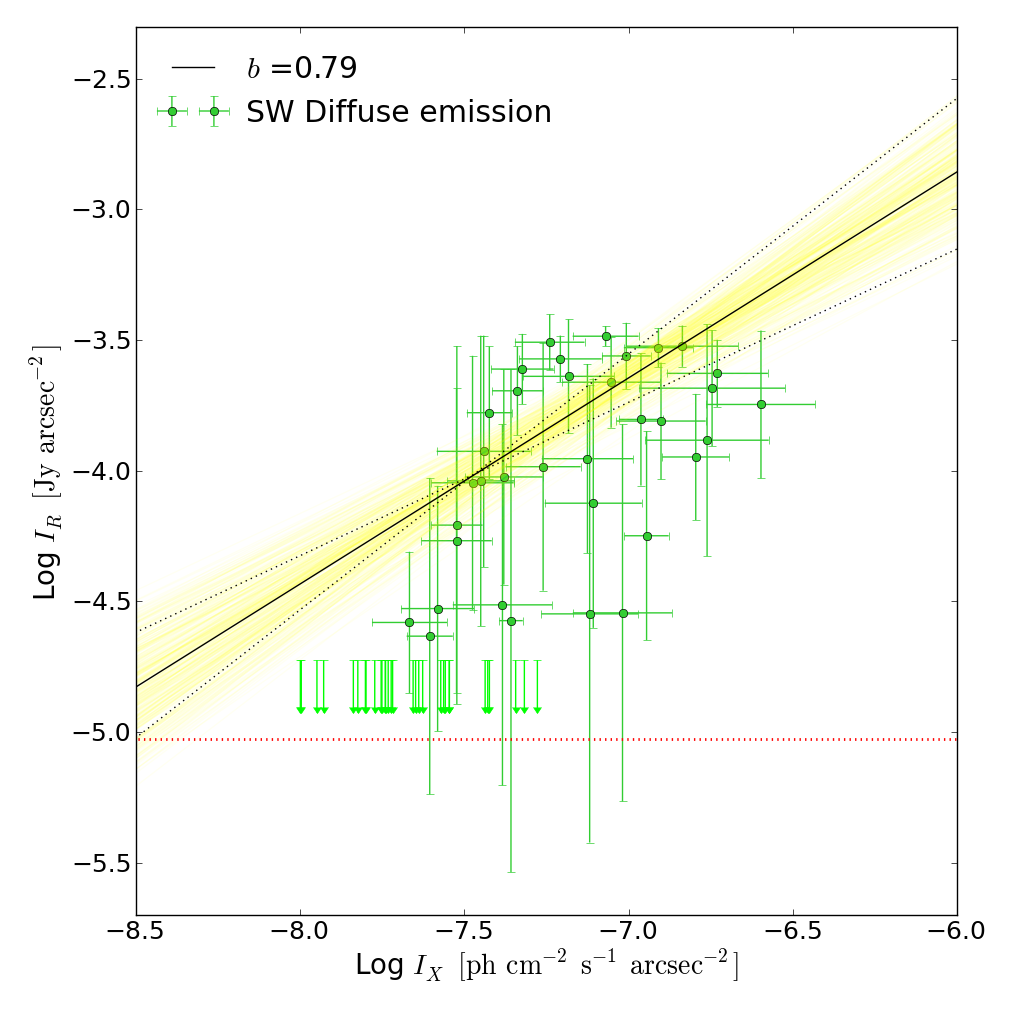}}
\caption{Radio-X-ray correlation for the diffuse emission on larger scale. \emph{Top:} LOFAR LBA image of RXJ1720.1, blanked outside a distance from the centre of r = 120 arcsec, at 15 arcsec resolution with contours at 3,10,20,30,50,60,100 $\times$ $\sigma$, where $\sigma=2.4\ \rm{mJy\ beam^{-1}}$. Overlaid is the grid in blue. The contribution of the mini halo and head-tail radio galaxy is excluded from the sampling (grey region). Green square indicates the region of south-west emission. \emph{Bottom left:} Radio-X-ray surface brightness correlation in the diffuse emission (blue points with associated statistical errors). The cyan arrows are the $2\sigma$ upper limits. The black line shows the best-fit correlation derived including upper limits, black dotted lines show the 25th and 75th percentile of the posterior distribution. The red dotted line marks $1\sigma$. \emph{Bottom right:} The same as before, but considering only the south-west diffuse emission (green points with associated statistical errors).}
\label{fig:Corr_Diff}
\end{figure*}

\section{Radio and X-ray comparison} \label{sec:comparison}
X-ray observations show that the cluster has a bright central core and on large scales it is relaxed with a regular morphology \citep[][]{Giacintucci2014}.
X-ray surface brightness and temperature profiles of the cluster revealed the presence of two cold fronts on opposite sides of the X-ray peak: at about 150 kpc to the south-east and at about 100 kpc north-west \citep[][]{Mazzotta2001,Mazzotta2008}.

In Fig. \ref{fig:X} we show a \emph{Chandra} X-ray image of the cluster in the $0.5 - 2.5$ keV band, obtained from the combination of three observations \citep[ObsIDs 1453, 3224, and 4631, for a total exposure of 42.5 ks, see][for details]{Mazzotta2008}. On the image we indicate with green lines the position of the cold fronts, and we overlaid the LOFAR LBA contours at resolution of 15 arcsec.
The diffuse radio emission extends outside the cluster core, as defined by the cold front boundaries, in the north-east south-west direction, perpendicular to the position of the cold fronts, and it lies in a region of lower X-ray surface brightness, with respect to the mini halo.
The X-ray image shows neither an excess of emission (i.e., substructures) nor a deficit (i.e., a cavity) in the region of the large-scale diffuse radio emission. 
We then check the residual image obtained after subtracting a 2D $\beta$-model \citep[][]{Cavaliere1976}, but also in this case there is no significant feature in correspondence of the LOFAR emission outside the cluster core.
However, a cavity could be present but not detectable due to low count rate outside the cluster core or due to projection effects.

\subsection{X-ray and radio surface brightness correlation} \label{sec:Ir-Ix}
To understand if there is an interplay between thermal and non-thermal emission inside and outside the cluster core, we performed a quantitative comparison between their surface brightness.
This connection would induce a spatial correlation between the radio ($I_{\rm R}$) and X-ray surface brightness ($I_{\rm X}$), which could be described by a power law in the log-log space:
\begin{equation}
    \log I_{\rm R} = b\log I_{\rm X} + c \ ,
\end{equation}
where a slope of  $b$ > 1 (super-linear relation) indicates that radio brightness declines faster than X-ray brightness or vice versa if $b$ < 1 (sub-linear relation).
The correlation is related to the particles origin \citep[][]{Dolag2000,Govoni2001,Brunetti2004,Pfrommer2008,ZuHone2013,ZuHone2015}. Hadronic models predict a super-linear relation, while turbulent re-acceleration scenarios can produce both super-linear and sub-linear relationships depending on the nature of the particles, their distribution in the cluster volume and on the scales considered.
A spatial correlation has been observed in several clusters.
In the case of giant radio haloes a sub-linear or linear scaling is generally found \citep[e.g.][]{Govoni2001,Botteon2020,Rajpurohit2021}, whereas mini haloes present a super-linear scaling between radio and X-rays \citep{Govoni2009,Ignesti2020}, suggesting an intrinsic difference in the nature of these radio sources.

We performed a point-to-point analysis of the emission in our source, considering its radio components separately. 
For each component, we constructed a square grid with cell sizes equal to the radio image FWHM and computed the mean surface brightness inside each cell.
To determine the best-fitting parameters of the
$I_{\rm R}$-$I_{\rm X}$ relation, we used the Linmix\footnote{https://linmix.readthedocs.io/en/latest/src/linmix.html} package \citep{Kelly2007}.
Linmix performs a Bayesian linear regression taking into account measurement uncertainties on both variables, intrinsic scatter, and upper limits. It calculates the posterior probability distribution of the model parameters, using a Markov Chain Monte Carlo (MCMC) method, and therefore it is accurate for both small and large sample sizes. We consider the mean of the posterior distribution as the best-fit parameters, and the scatter of the 25th and 75th percentile with respect to the mean as an estimate of the slope uncertainty. The correlation strength was measured by using the Pearson and Spearman correlation coefficients.

For the study of the central emission we used a LOFAR HBA image at 6 arcsec resolution. The \emph{Chandra} X-ray image was smoothed to the same resolution.
We created a grid at the location of the mini halo emission, indicated with red squares in the left panel of Fig. \ref{fig:Corr_MH}, and on the east extension, yellow squares.
The right panel of Fig. \ref{fig:Corr_MH} shows the point-to-point comparison between the X-ray and radio surface brightness in log-log scale. Each point represents the brightness in each cell of the grid with associated statistical errors (red points for mini halo and orange points for east extension).
We note that the east extension and the mini halo follows different trends, suggesting the two radio emission are different.
We then computed the mini halo correlation considering only the red points.
In the plot the solid black line represents the best fit of $b$, and the dotted black lines the 25th and 75th percentile. We list the results of the fit in Table \ref{tab:corr}.
We find that the values of $I_{\rm R}$ and $I_{\rm X}$ are strongly positively correlated (correlation coefficients near +1) and the slope is super-linear, $b = 1.78\pm0.22$. 
This implies that the non-thermal and thermal plasma in the mini halo region are connected, and suggests a peaked distribution of relativistic electrons and magnetic field in the cluster core. 
We note that \cite{Ignesti2020} have already computed a radio X-ray correlation for this source, using GMRT data at 612 MHz, and they found a shallower slope of $b=1.5\pm0.1$. Several factors may be the cause of this difference: they use a different method to estimate the parameters (BCES) and they compute the surface brightness from a different grid, considering all the radio emission visible at 612 MHz, so including the east extension, and excluding a wider region around the bright central galaxy.

We performed the same analysis for the large-scale diffuse emission. This time we used the LBA image, in which the diffuse emission is more evident, with a resolution of 15 arcsec and we smoothed the X-ray image to the same resolution.
We constructed a grid with cells as large as the beam, covering all the emission inside a circle of 120 kpc, from which we excluded areas containing the mini halo, east extension and head-tail radio galaxy emission. In Fig. \ref{fig:Corr_Diff} we show on the top the grid used overlaid on the radio image, while on the bottom left panel we plot the data together with the best fit line of $b$ (black solid line), and the 25th and 75th percentile lines (black dotted lines).
Radio values below $2\sigma_{\rm rms}$ are considered as upper limits and included in the fit.
We show the results of the fit in Table \ref{tab:corr}.
In this case we found a moderate positive correlation (correlation coefficients around 0.7) and a slope $b=1.00\pm0.11$. If instead we consider only the south-west part of the diffuse emission (indicated by a green square in Fig. \ref{fig:Corr_Diff}, top panel), which is less contaminated by other components, the correlation is sub-linear, with a slope $b=0.79\pm0.20$ (see Fig. \ref{fig:Corr_Diff}, bottom right panel).
We also compute the correlation using the HBA image with a resolution of 15 arcsec, to make a comparison with the results obtained with the LBA image. We found a slope $b = 1.10\pm0.14$ ($b=0.87\pm0.25$ for the SW diffuse emission) which is comparable to that obtained at lower frequency, but in this case we obtained a weaker correlation, with correlation coefficients $\sim0.6$  ($\sim0.7$ for the SW diffuse emission).
We therefore found for the diffuse emission a less steep correlation, compared to that found for the mini halo, and closer to the values found in the literature for giant radio haloes.

We note that using cells with the same size of the radio image FWHM could generate biases because contiguous cells are not statistically independent. This is especially true when the emission is sampled with a small number of cells ($\sim 20-30$). To account for this, we repeat the analysis increasing slighter the size of the cells for both mini halo and diffuse emission, founding comparable results. When the size is doubled, on the other hand, we do not have enough points to well constrain the correlation. 
Another approach to the problem was proposed by \cite{Ignesti2020}, with a sampling of the emission of mini haloes through non-fixed grids and Monte Carlo analysis. We note that the dispersion of the $b$ values they get in their sample is around 0.14. The slope uncertainties reported in Table \ref{tab:corr} are therefore wide enough to take this effect into account.

We can therefore confidently say that the mini halo has a super-linear correlation, while the diffuse emission on a larger scale follows a linear/sub-linear trend, suggesting an intrinsic difference in their thermal properties.

\begin{table}
    \centering
    \caption{$I_{\rm R}$-$I_{\rm X}$ correlation.}
    \renewcommand\arraystretch{1.2}
    \begin{tabular}{ccccc} \hline
         Region &$\nu$ &$b$ &$r_p$$^1$ &$r_s$$^2$\\\hline
        Mini halo &144 MHz &$1.78\pm0.22$ &0.96 & 0.94 \\
        East extension &144 MHz &$0.61\pm0.08$ &0.79 & 0.80 \\
        Diffuse emission &54 MHz &$1.00\pm0.11$ &0.71 &0.67 \\
        &144 MHz &$1.10\pm0.14$ &0.61 &0.58\\
        SW Diffuse emission &54 MHz &$0.79\pm0.20$ &0.76 &0.78 \\
        &144 MHz &$0.87\pm0.25$ &0.72 &0.67\\
         \hline \end{tabular}
         \\ Notes. $1$: Pearson correlation coefficient, $2$: Spearman correlation coefficient.
        \label{tab:corr} 
\end{table}

\subsection{X-ray surface brightness vs spectral index}
\begin{figure*}
\subfloat{
\centering
\includegraphics[width=0.45\textwidth]{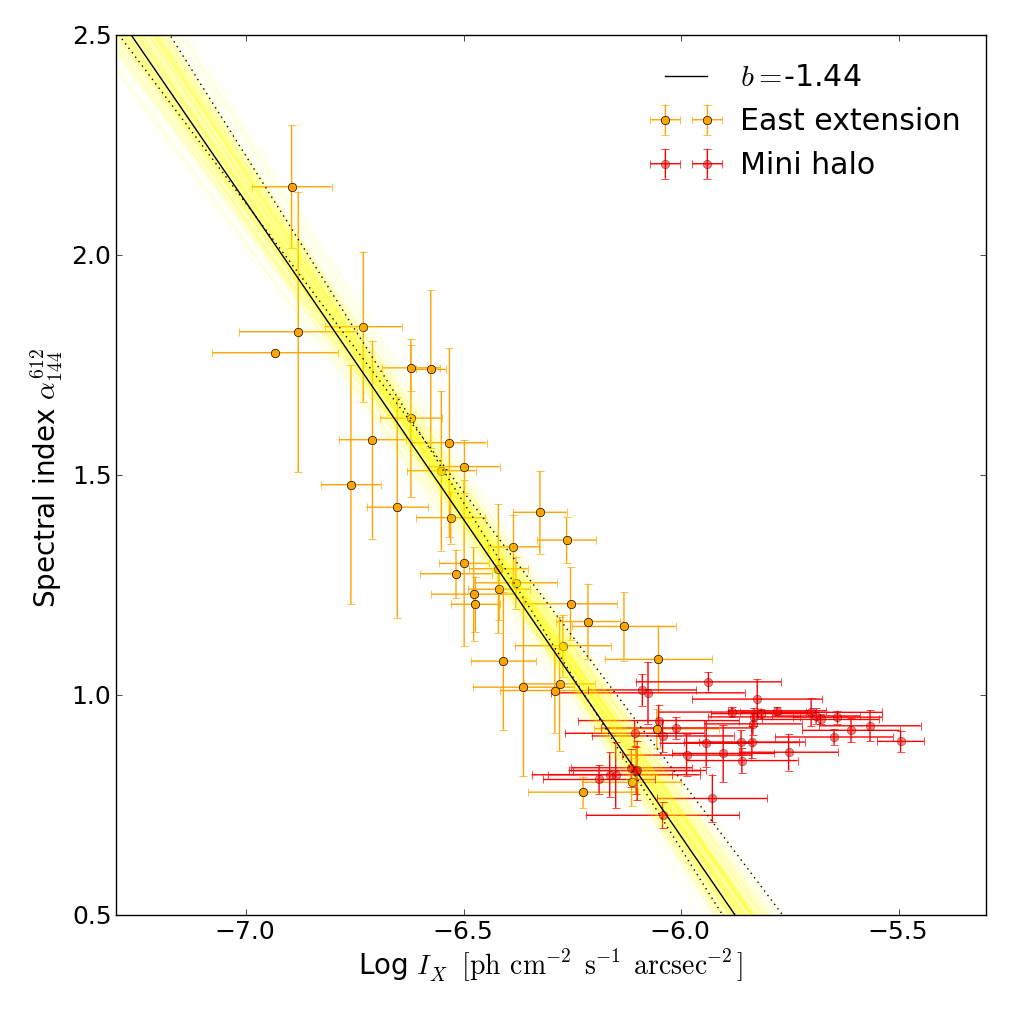}}
\qquad
\subfloat{
\includegraphics[width=0.45\textwidth]{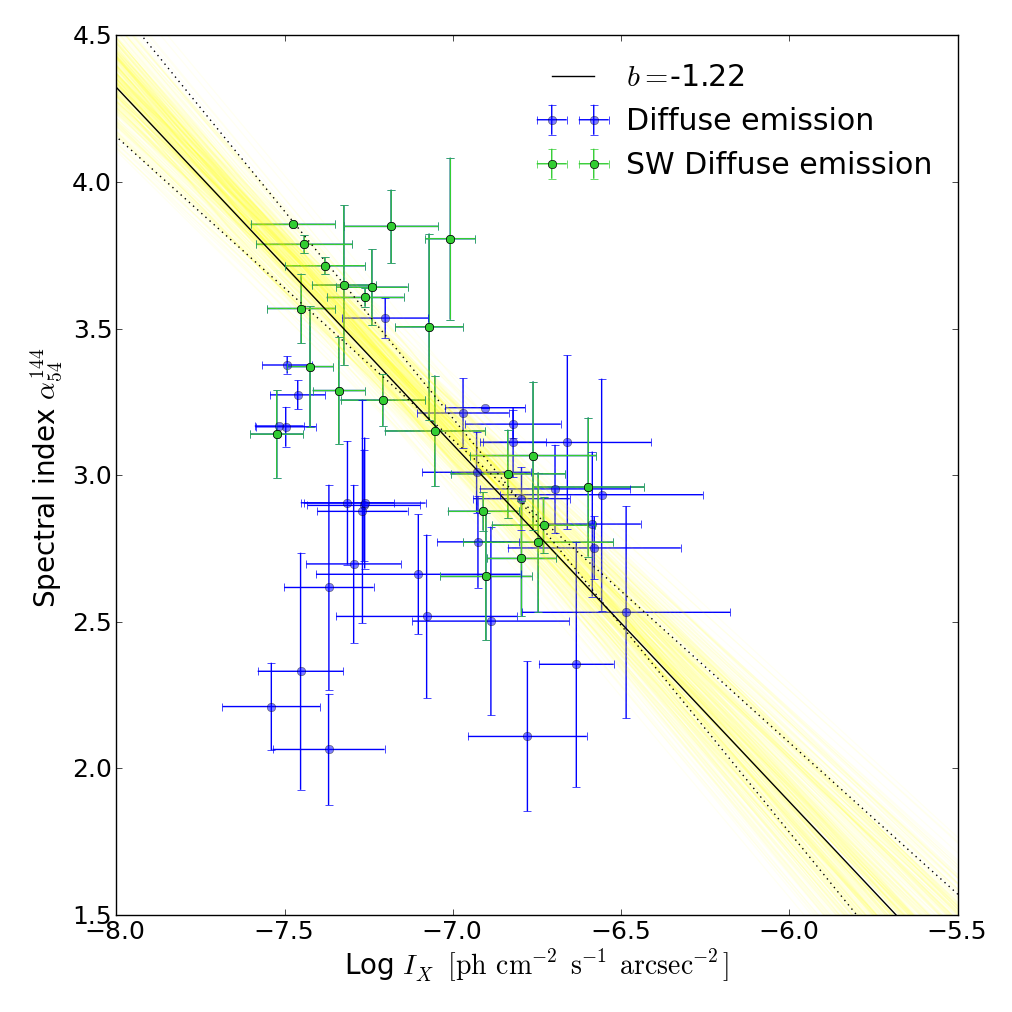}}
\caption{\emph{Left:} Spectral index - X-ray surface brightness correlation of the central emission. Orange points, with associated statistical errors, are from the east extension, while red points from the mini halo. The black line shows the best-fit correlation on east extension, black dotted lines show the 25th and 75th percentile of the posterior distribution. \emph{Right:} Spectral index - X-ray surface brightness correlation of the diffuse emission on larger scale (blue points, with associated statistical errors). The points related to the SW part of the diffuse emission are indicated in green. The black line shows the best-fit correlation, black dotted lines show the 25th and 75th percentile of the posterior distribution.}
\label{fig:corr_spix}
\end{figure*}

We also studied the point-to-point distribution of the radio spectral index in relation to the thermal emission.
For the study of the central radio emission we used the high-frequency spectral index map, between $144 - 612$ MHz, for which the resolution is higher. For the diffuse emission on larger scale, instead, we used the low-frequency spectral index map, between $54 - 144$ MHz.
To extract the X-ray surface brightness and spectral indices, we used the same grid regions presented in Sec. \ref{sec:Ir-Ix}, considering the radio emission above $3\sigma$ level.
The results are shown in Fig. \ref{fig:corr_spix}, left panel for the central emission and right panel for diffuse emission on larger scale.
There is evidence of an anti-correlation for the east extension (orange points), while for the mini halo radio spectral index and X-ray surface brightness are not correlated (red points). 
An anti-correlation is also present for the south-west part of the diffuse emission (green points). If, on the other hand, all the diffuse emission is considered (blue points), the scatter of the points is greater and they appear uncorrelated, suggesting a contamination of other components to this emission.
To compute the significance of the correlation, we estimate the Spearman and Pearson correlation coefficients and fit the data with a relation of the form:
\begin{equation}
    \alpha = b \log I_{\rm X} + c .
\end{equation}
As in the previous section, we performed a linear regression using Linmix.
The resultant parameters are reported in Table \ref{tab:corr2}.
There is a moderate negative correlation between the spectral index and X-ray surface brightness of the two components with a super-linear best-fit slope, indicating that the spectral index is flatter at high X-ray surface brightness and steepens in low X-ray brightness regions.
Recently, an anti-correlation between these two quantities has been observed for the radio halo in MACS J0717.5+3745 \citep[][]{Rajpurohit2021}, conversely for the halo in Abell 2255 \cite{Botteon2020} found a positive correlation.

\begin{figure*}
\subfloat{
\centering
\includegraphics[width=0.45\textwidth]{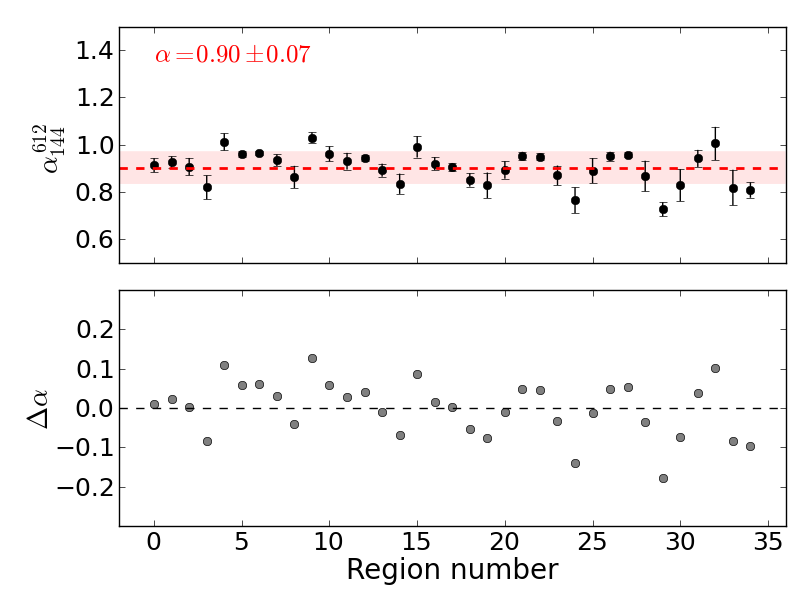}}
\qquad
\subfloat{
\includegraphics[width=0.45\textwidth]{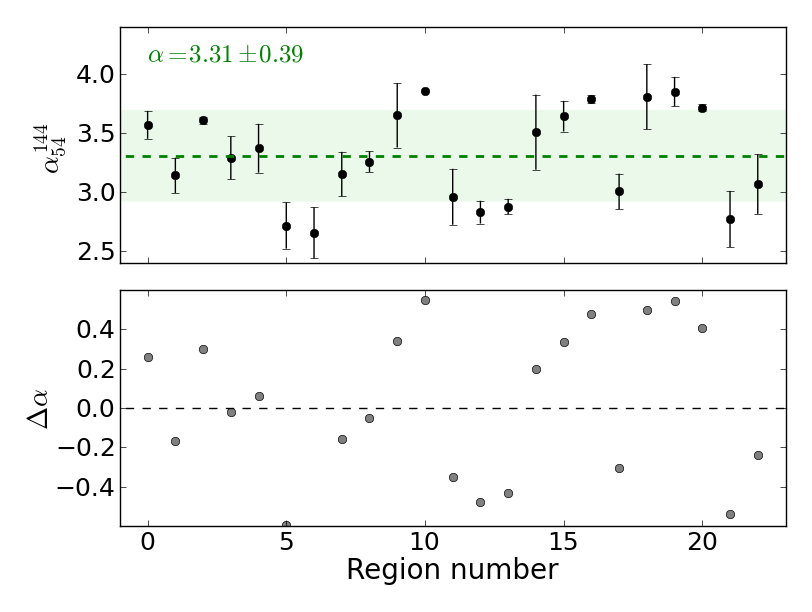}}
\caption{Spectral index distribution across the mini halo (\emph{Left}, $\alpha_{144}^{612}$) and across the SW diffuse emission (\emph{Right}, $\alpha_{54}^{144}$). The spectral indices were extracted in regions indicated in Fig. \ref{fig:Corr_MH} (red squares) and in Fig. \ref{fig:Corr_Diff} (green square), respectively. The dashed horizontal line indicates the mean spectral index, while the coloured region represents the standard deviation. The lower panels show the residuals of $\alpha$ with respect to the mean spectral index.}
\label{fig:spix_dist}
\end{figure*}

In Fig. \ref{fig:spix_dist} we show the spectral index distribution across the mini halo (left panel) and the SW diffuse emission (right panel).
On mini halo scale there is a small scatter in the spectral index distribution ($\Delta\alpha\sim 0.1$), while a large scatter is observed in the SW diffuse emission on larger scales, with values in the range $2.6 \le \alpha \le 3.9$.

\begin{table}
    \centering
    \caption{$\alpha$-$I_{\rm X}$ correlation.}
    \renewcommand\arraystretch{1.2}
    \begin{tabular}{ccccc} \hline
         Region &$\nu$ &$b$ &$r_p$$^1$ &$r_s$$^2$\\\hline
         Mini halo &144-612 MHz &$0.18\pm0.07$ &036 & 0.42 \\
        East extension &144-612 MHz &$-1.44\pm0.14$ &-0.86 &-0.87 \\
        Diffuse emission & 54-144 MHz &$-0.62\pm0.16$ & -0.32 & -0.34 \\
        SW Diffuse emission &54-144 MHz &$-1.22\pm0.20$ &-0.71 &-0.67 \\
         \hline \end{tabular}
         \\ Notes. $1$: Pearson correlation coefficient, $2$: Spearman correlation coefficient. 
        \label{tab:corr2} 
\end{table}

\section{Discussion} \label{sec:discussion}
In this work we presented new LOFAR LBA observations of the galaxy cluster RXJ1720.1, which presents a central mini halo and fainter diffuse emission on larger scales. 
We performed spectral analysis of the source and a comparison between radio and X-ray properties to try to understand the origin of this hybrid morphology.
Our results reveal that the mini halo and the more diffuse emission show distinct features, suggesting the radio emission inside and outside the cluster core have a different origin.
They could be either generated by different mechanisms, such as hadronic origin for the mini halo and turbulence for the more extended emission, either by the same mechanism, such as turbulence, with different origins or the same origin but in different conditions of e.g. magnetic field or distribution of seed electrons. Another possibility is that different micro-physical conditions of the ICM in the cluster volume can change the acceleration rate of the particles \citep[][]{Brunetti2011}.

\subsection{Central emission}
The central emission of RXJ1720.1 consists of a bright central mini halo and a fainter emission extending to the east, delimited by two cold fronts. 
The east extension was previously interpreted as a continuation of the mini halo \citep[][]{Giacintucci2014}. 
However, our results showed that the mini halo and the east extension present different trend for the $I_{\rm R}-I_{\rm X}$ and $\alpha-I_{\rm X}$ correlations. So we decided to exclude this emission from the analyses concerning the mini halo. The east extension on the other hand follows correlations that are more similar to those of the SW diffuse emission. This could indicate that also the east extension is powered by the same mechanism responsible for the SW diffuse emission. In any case, because of the nearby AGN tail, it is likely that we are observing the superposition of the halo with the tail. As these emissions cannot be separated, we focused our analysis on the SW diffuse emission only (see Section \ref{Diff}).

The mini halo has a mean spectral index of $\alpha_{144}^{612}=0.93\pm0.10$ with little spectral index variations.
It follows a super-linear correlation between radio and X-ray surface brightness, indicating that the relativistic electrons and magnetic field are more concentrated around the central AGN. The relativistic particles injected by the central AGN can then play a role both directly generating secondary electrons or as seed particles re-accelerated by turbulence.

\begin{figure}
\centering
\includegraphics[width=0.5\textwidth]{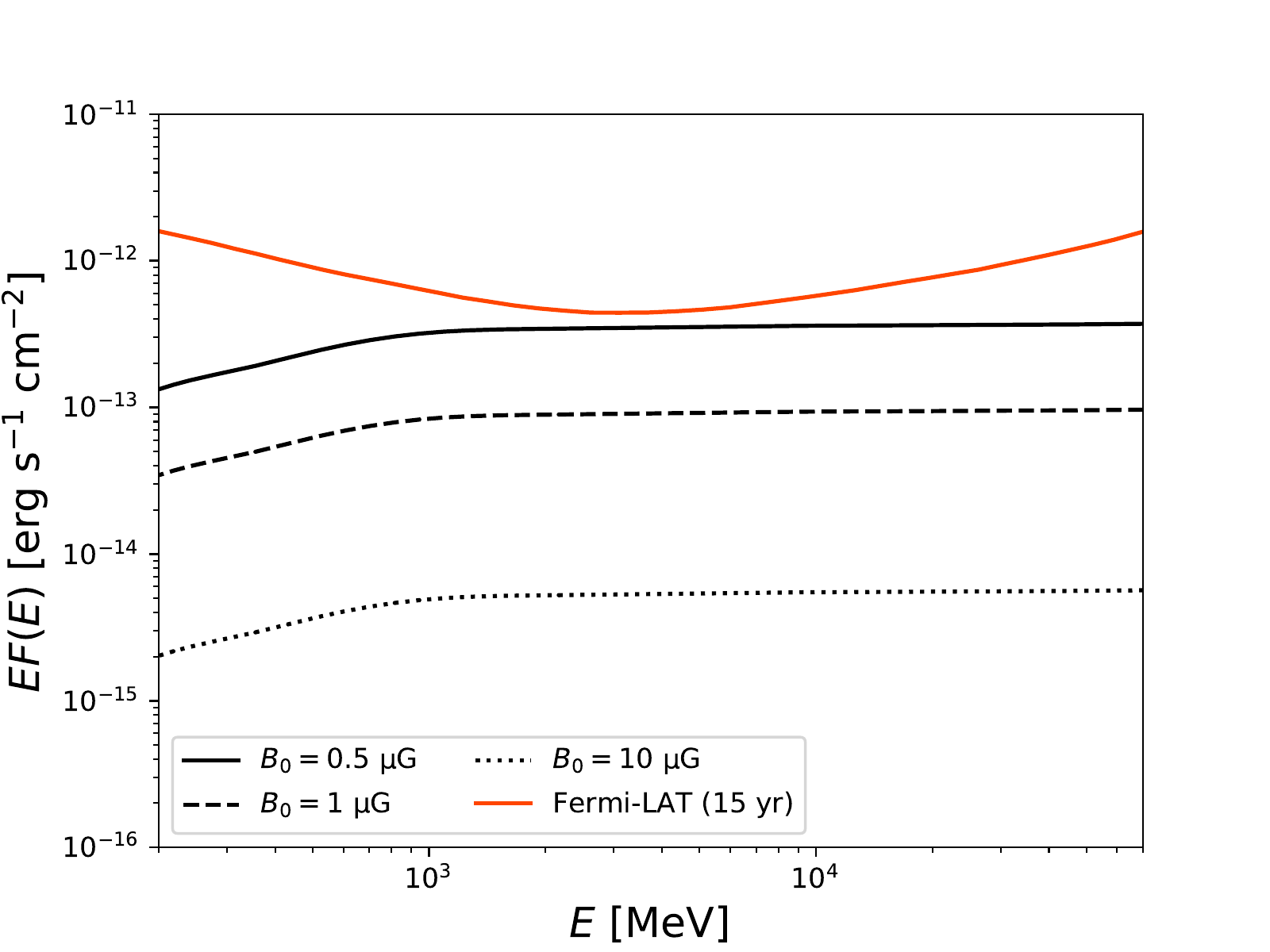}
\caption{$\gamma$-rays spectrum for the mini halo considering a central magnetic field of $B_0 = 0.5\ \rm{\mu G}$ (black solid line), $B_0 = 1\ \rm{\mu G}$ (black dashed line) and of $B_0 = 10\ \rm{\mu G}$ (black dotted line) compared with the \emph{Fermi}-LAT 15 yrs detection limit (orange line).}
\label{fig:gamma}
\end{figure}

A test of hadronic models \citep[][]{Pfrommer2004,Ensslin2011,ZuHone2015} for mini haloes can be obtained from the $\gamma$-rays due to the decay of $\pi^0$ that are produced by the same chain of secondary electrons.
\cite{Ignesti2020} calculated the $\gamma$-ray emission of a sample of mini haloes. Their predicted fluxes are below the \emph{Fermi}-LAT \footnote{Large Area Telescope (LAT) onboard of the \emph{Fermi} satellite} detection limit, hence the hadronic model does not violate the current non-detection of diffuse $\gamma$-ray emission.
We use the formalism in \cite{Brunetti2017} to calculate the expected $\gamma$-ray flux from the mini halo region of RXJ1720.1 assuming a pure hadronic origin of the mini halo. The expected $\gamma$-rays are a factor 10 below \emph{Fermi}-LAT sensitivity (assuming 15 yrs of data) assuming an average magnetic field of $B_0 = 1\ \rm{\mu G}$ in the core of the cluster (Fig. \ref{fig:gamma}). Larger $\gamma$-ray fluxes are predicted for weaker magnetic fields, yet we find that for $B_0 \leq 0.5\ \mu$G the hadronic model predicts $\gamma$-rays below the current \emph{Fermi}-LAT sensitivity. The $\gamma$-ray flux would be even lower in the case cosmic-ray protons and their secondaries are re-accelerated by turbulence in the core \citep{Brunetti2011,Pinzke2017}.

In conclusion, although in previous literature a re-acceleration of seeds electrons origin of the mini halo in RXJ1720.1 was favoured \citep{Giacintucci2014,ZuHone2013}, we found that also a hadronic origin may provide a valid interpretation.

\subsection{Diffuse emission outside the cluster core}\label{Diff}
Low-frequency observations with LOFAR have revealed the presence of fainter diffuse emission outside the cluster core of RXJ1720.1.
This emission extends in the north-east south-west direction, reaching a total extension of 600 kpc, and has an ultra-steep spectrum, with a mean spectral index of $\alpha_{54}^{144}=3.2\pm0.2$. 
The X-ray emission instead has a regular morphology on large scales.
We propose two possible scenarios:
\begin{itemize}
    \item The cluster underwent a minor merger that injected turbulence into the medium and re-accelerated particles on large scales. In this case the large-scale diffuse emission is similar to a giant radio halo.   
In fact it extends up to 600 kpc and there is evidence of steep emission even in the central regions of the cluster. Moreover, we found a modest linear/sub-linear correlation between radio and X-ray surface brightness. 
This emission shows an ultra-steep spectrum $\alpha_{54}^{144}=3.2\pm0.2$, as observed in rare cases of very gentle particle re-acceleration \citep[][]{deGasperin2017,Hodgson2021}.
In our source, the turbulence could be generated as a consequence of a minor/off-axis merger.
Steep-spectrum diffuse emission is indeed predicted by re-acceleration models in connection with less energetic merger events \citep[e.g.][]{Cassano2006,Brunetti2008,Wilber2018}.
The minor merger has then dissipated enough energy into the ICM to accelerate particles in a large volume, without disrupting the cool-core of the cluster. In this scenario, therefore, both a mini halo and a giant radio halo coexist, as proposed for the cool-core cluster PSZ1G139.61+24.20 \citep[][]{Savini2018}.
The large-scale sloshing should have left an imprint also in the distribution of the thermal emission.
However, the X-ray map of RXJ1720.1 does not show an excess of surface brightness at the location of the SW diffuse radio emission.
Evidence in support of this scenario is the detection of group-scale substructures with optical spectroscopy \citep[][]{Owers2011}, that may have perturbed the medium and produced the emission we observe in the radio band. The most likely perturber is a substructure located south-west of cluster centre at a distance of $\sim400$ kpc, in the same direction of the large-scale diffuse radio emission we observed. A slight increase observed in the weak lensing map of \cite{Okabe2010} supports this hypothesis.

\item The radio emission located south-west of the cluster core is the relic of a giant bubble generated by the jet activity of the central AGN.
The very steep spectrum we found for this emission suggests that the bubble is very old. The radiative age of a bubble can be calculated from the frequency at which the spectrum of a synchrotron-emitting population, initially described by a power-law with a spectral index between $0.6-0.8$, steepens due to energy losses. Assuming an initial power-law of $\alpha=0.7$ and an equipartition magnetic field of $B_{\rm eq}\sim5.7\rm{\mu G}$, we estimated a bubble age of $t\sim 350$ Myr.
The counter-part of this bubble is located north-east of the centre, where we also found evidence of steep emission between the mini halo and the head-tail radio galaxy.
This scenario is supported by the recent discovery of a giant X-ray cavity in the Ophiucus galaxy cluster \citep[][]{Werner2016,Giacintucci2020}, filled by radio emission with a steep spectrum ($\alpha=2.4\pm0.1$).
In our source no X-ray cavity has been identified. However, it could be present but not detectable due to small contrast in the X-ray image outside the cluster core or due to projection effects.
Other results against this scenario are: the detection of ultra-steep emission in the mini halo region, which suggests a spherical morphology like a halo rather than two bubbles; the radio and X-ray surface brightness are positively correlated, not expected in the presence of a bubble.
\end{itemize}
In conclusion, we provide evidence that the large-scale diffuse emission should be generated by re-acceleration of particles after a minor/off-axis merger.


\section{Conclusions} \label{sec:conclusions}
In this paper, we have presented new LOFAR LBA observations of the source RX J1720.1+2638. These data, combined with archival data at higher frequencies, allow us to perform a resolved spectral study and investigate the possible origins of this peculiar source.
A comparison between radio and X-ray properties completes this work. The main results are summarized below:
\begin{itemize}
\item RXJ1720.1 is a cool-core cluster with an hybrid morphology in the radio band, it is composed of a central bright mini halo and fainter diffuse emission, only detect with LOFAR, extending outside the cluster core in the north-east south-west direction.
The mini halo and the diffuse emission on larger scales show different features, suggesting they have a different nature.

\item The mini halo has a spectral index of $\alpha_{144}^{612}=0.93\pm0.10$, the correlation between radio and X-ray surface brightness is super-linear and the $\gamma$-ray fluxes we derived do not violate \emph{Fermi}-LAT detection limits.  
Our results indicate that the mini halo could have a hadronic origin, as a valid alternative to the re-acceleration models previously proposed in the literature.

\item For the first time we estimate the spectral index of the diffuse emission outside the cluster core, finding an ultra-steep spectrum, with a mean spectral index of $\alpha_{54}^{144}=3.2\pm0.2$.

\item There is neither an excess or a deficit of X-ray emission in the region of the large-scale diffuse emission, but we found a positive correlation between radio and X-ray surface brightness.

\item We argue that the large-scale diffuse emission was generated by re-acceleration of particles after a minor merger. We also speculated that it could be a relic bubble, but there is no observational evidence to support this hypothesis.
\end{itemize}
 
 Our work shows that LBA observations are of great importance to constrain the spectral properties of diffuse radio emission observed in clusters of galaxies. There are few other sources identified so far that exhibit a hybrid morphology such as RXJ1720.1. Low frequency radio observations of these sources would allow to conduct a spectral study of a small statistical sample and therefore to verify the hypotheses proposed in this paper for the origin of the large-scale radio emission.

\section*{Acknowledgements}
NB, AB, CJR acknowledge support from the ERC through the grant ERC-
Stg DRANOEL n. 714245.
HE acknowledges support from the DFG, Germany [427771150]. 
Basic research in radio astronomy at the Naval Research Laboratory is supported by 6.1 Base funding.
AB acknowledges support from the VIDI research programme with project number 639.042.729, which is financed by the Netherlands Organisation for Scientific Research (NWO).
GB, RC, FG and MR acknowledge support from INAF mainstream project "Galaxy Clusters science with LOFAR".
AD acknowledges support by the BMBF Verbundforschung under the grant 05A20STA.
ACE acknowledges support from STFC grant ST/P00541/1.
RJvW acknowledges support from the ERC Starting Grant ClusterWeb 804208.  
LOFAR, the Low Frequency Array designed and constructed by ASTRON (Netherlands Institute for Radio Astronomy), has facilities in several countries, that are owned by various parties (each with their own funding sources), and that are collectively operated by the International LOFAR Telescope (ILT) foundation under a joint scientific policy.
This research made use of the LOFAR IT computing infrastructure supported and operated by INAF, and by the Physics Dept. of Turin University (under the agreement with Consorzio Interuniversitario per la Fisica Spaziale) at the C3S Supercomputing Centre, Italy.
The J\"ulich LOFAR Long Term Archive and the German LOFAR network are both coordinated and operated by the J\"ulich Supercomputing Centre (JSC), and computing resources on the supercomputer JUWELS at JSC were provided by the Gauss Centre for supercomputing e.V. (grant CHTB00) through the John von Neumann Institute for Computing (NIC). 
The scientific results reported in this article are based in part on data obtained from the VLA Data Archive, the GMRT Data Archive and the \emph{Chandra} Data Archive. 
This research made use of the following Python packages: APLpy \citep{Robitaille2012}, Astropy \citep{Astropy2013} and NumPy \citep{vanderwalt2011}.
 
\section*{Data availability}
The data underlying this article will be shared on reasonable request to the corresponding author.

\bibliographystyle{mnras}
\bibliography{biblio}

\bsp	
\label{lastpage}
\end{document}